\documentclass[prb,twocolumn]{revtex4-2}
\usepackage{graphicx}
\usepackage{amsmath,amsfonts,amssymb}
\usepackage{bm}
\usepackage{hyperref}
\usepackage{physics}
\usepackage{mathtools}
\usepackage{booktabs}

\begin{document}

\title{Continuous crossover between high-pressure ice phases VII and X driven by monopole screening: a model study}

\author{Sena Watanabe}
\email{watanabe-sena397@g.ecc.u-tokyo.ac.jp}
\affiliation{Department of Applied Physics, The University of Tokyo, Tokyo 113-8656, Japan}

\author{Yukitoshi Motome}
\email{motome@ap.t.u-tokyo.ac.jp}
\affiliation{Department of Applied Physics, The University of Tokyo, Tokyo 113-8656, Japan}

\author{Haruki Watanabe}
\email{hwatanabe@ust.hk}
\affiliation{Department of Physics, Hong Kong University of Science and Technology, Clear Water Bay, Hong Kong, China}
\affiliation{Institute for Advanced Study, Hong Kong University of Science and Technology, Clear Water Bay, Hong Kong, China}
\affiliation{Center for Theoretical Condensed Matter Physics, Hong Kong University of Science and Technology, Clear Water Bay, Hong Kong, China}
\affiliation{Department of Applied Physics, The University of Tokyo, Tokyo 113-8656, Japan}

\begin{abstract}
The proton-disordered molecular phase of water ice (ice-VII) and its ultrahigh-pressure non-molecular phase (ice-X) share identical macroscopic crystal symmetry (space group $Pn\bar{3}m$). 
This raises a fundamental thermodynamic question: are they distinct phases separated by a singularity, or are they adiabatically connected via a continuous crossover? 
To resolve this paradox, we investigate the finite-temperature phase diagram of high-pressure ices VII and X, as well as VIII, the proton-ordered phase that emerges at lower temperatures, using an effective classical spin-$1$ Blume-Capel model on the pyrochlore lattice. 
Through Monte Carlo simulations, we demonstrate that within this model, the transformation between the states corresponding to ice-VII and ice-X lacks a thermodynamic singularity, as characterized by non-divergent and non-coinciding peaks in the specific heat and susceptibility associated with the $S^z=0$ occupation. 
We attribute this continuous crossover behavior to the topological fragility of the hydrogen-bond network: the thermal proliferation of point-like monopole excitations (violations of the ice rules) induces Debye-H\"{u}ckel screening of the emergent gauge field, destroying the topological Coulomb phase at any finite temperature. In contrast, the destruction of the proton-ordered ice-VIII phase involves spontaneous symmetry breaking and remains a first-order phase transition. Our findings provide a microscopic rationale that reconciles the macroscopic crystallographic symmetries of dense ice with its underlying topological properties.
\end{abstract}

\maketitle

\section{Introduction}
\label{sec:intro}
The phase diagram of water is arguably the first and most familiar phase diagram encountered in the study of physics and chemistry. In its standard classical form, the solid (ice), liquid (water), and gaseous (vapor) phases are separated by lines of first-order phase transitions. However, a fundamental thermodynamic concept is embedded in the liquid-gas boundary: it terminates at a critical point~\cite{Stanley1971}. Beyond this point, the liquid and gaseous states can be continuously transformed into one another by appropriately tuning temperature and pressure without crossing a thermodynamic singularity. From the perspective of adiabatic continuity, the liquid and gas phases lack a phase boundary and ultimately belong to the same fluid phase~\cite{Chaikin1995}.

In contrast, the solid phase is separated from both the liquid and gaseous phases. This thermodynamic boundary is dictated by the spontaneous breaking of continuous translational and rotational symmetries inherent in crystallization. The classification of macroscopic phases based on spontaneous symmetry breaking and the emergence of local order parameters has served as the cornerstone of condensed matter physics, known as the Landau-Ginzburg paradigm~\cite{Landau1937, Anderson1984}.

In recent decades, condensed matter physics has evolved beyond the Landau paradigm, revealing that phases sharing identical symmetries can still be distinct due to their underlying topological properties. This realization has led to the classification of symmetry-protected topological (SPT) phases~\cite{Chen2012, Senthil2015} and topologically ordered phases characterized by long-range quantum entanglement~\cite{Wen1990, Kitaev2006}. However, a caveat remains: most of these topological distinctions are defined primarily at absolute zero ($T=0$). Once finite temperatures are introduced, thermal fluctuations and the proliferation of fractionalized topological excitations (e.g., anyons) generally destroy this topological protection~\cite{Nussinov2008, Hastings2011}. Consequently, topological phase transitions---those occurring without spontaneous symmetry breaking---rarely manifest in macroscopic, finite-temperature phase diagrams, leaving the symmetry-breaking paradigm largely dominant for understanding thermal phase transitions.

Returning to the phase diagram of water, the solid region itself is extraordinarily rich and complex. To date, approximately twenty distinct crystalline polymorphs of ice have been discovered or predicted across a vast range of thermodynamic conditions, extending from extreme high pressures to the negative-pressure regime~\cite{RevModPhys.84.885, Salzmann2011, doi:10.1126/sciadv.1501010, Salzmann2019, Hansen2021}.
Most of these diverse ice phases can be distinguished within the Landau-Ginzburg paradigm by their different crystallographic space groups, which represent distinct patterns of spatial symmetry breaking.

However, an apparent paradox emerges at extremely high pressures. The room-temperature high-pressure phase, ice-VII, and the ultrahigh-pressure non-molecular phase, ice-X, are experimentally reported and theoretically predicted to share the exact same macroscopic space group symmetry ($Pn\bar{3}m$)~\cite{Holzapfel1972, Kuhs1984, Hemley1987}. In ice-VII, protons are dynamically disordered, hopping between two equivalent off-center sites along the oxygen-oxygen bonds, satisfying the Bernal-Fowler ice rules~\cite{Bernal1933}. In ice-X, the hydrogen bonds symmetrize, and the protons become localized at the midpoints of the bonds. 
Because no macroscopic symmetry is broken during this structural evolution, a fundamental question arises: are ice-VII and ice-X distinct thermodynamic phases separated by a singularity, or are they---much like liquid water and vapor---fundamentally the same phase connected by a continuous crossover?

To address this question from a statistical mechanics perspective, we investigate a unified effective spin-$1$ model defined on the pyrochlore lattice. In this model, a single-ion anisotropy parameter $\Delta$ acts as a chemical potential that penalizes asymmetric proton configurations, driving the system toward the centered-proton ice-X phase.

Recently, Pandey and Damle~\cite{Pandey2025} investigated the pure nearest-neighbor version of this $S=1$ model in the strict low-temperature limit ($T/J \to 0$), where thermal monopole excitations (violations of the ice rules) are completely suppressed. In this monopole-free limit, they demonstrated the existence of three distinct low-temperature phases---a short-range correlated paramagnet and two topologically distinct Coulomb liquids---separated by sharp phase transitions. However, in physical ice crystals at finite temperatures, thermal point-like monopoles are thermodynamically inevitable. 
The present authors investigated this issue through duality mappings to 3D $XY$ and Ising loop-gas models~\cite{Watanabe2026}.
They demonstrated that the topological phase boundaries present in the monopole-free limit degrade into continuous crossovers at any finite temperature. Specifically, thermal monopoles act as a symmetry-breaking field in the continuous $XY$ picture and topologically sever defect strings in the Ising loop-gas picture.

Building upon this analytical foundation, the present study employs Monte Carlo simulations to investigate the full phase diagram incorporating a finite next-nearest-neighbor interaction~\cite{Ishizuka2014}, which is essential for capturing the thermodynamic competition involving the macroscopic polarization (represented by a ferromagnetic order in the spin model) corresponding to the low-temperature proton-ordered ice-VIII phase. 
We demonstrate that, due to the thermal proliferation of point-like monopole excitations, the finite-temperature transformation between the model states corresponding to ice-VII and ice-X does not exhibit a thermodynamic singularity, confirming its nature as a continuous crossover.

The remainder of this paper is organized as follows. In Sec.~\ref{sec:background}, we review the crystal structures of high-pressure ice, summarize the experimental and theoretical background concerning their phase transitions, and discuss the topological conditions for finite-temperature phase transitions without spontaneous symmetry breaking. In Sec.~\ref{sec:model}, we formulate the effective microscopic spin-$1$ model. 
In Sec.~\ref{sec:methods}, we define the relevant physical quantities and detail our Monte Carlo methods.
In Sec.~\ref{sec:results}, we present the Monte Carlo simulation results. Finally, in Sec.~\ref{sec:discussion}, we discuss the thermodynamic nature of the crossovers in light of monopole screening and conclude our findings.

\section{Background}
\label{sec:background}

\subsection{Crystal structures}

\begin{figure}[t]
  \centering
  \includegraphics[width=\linewidth]{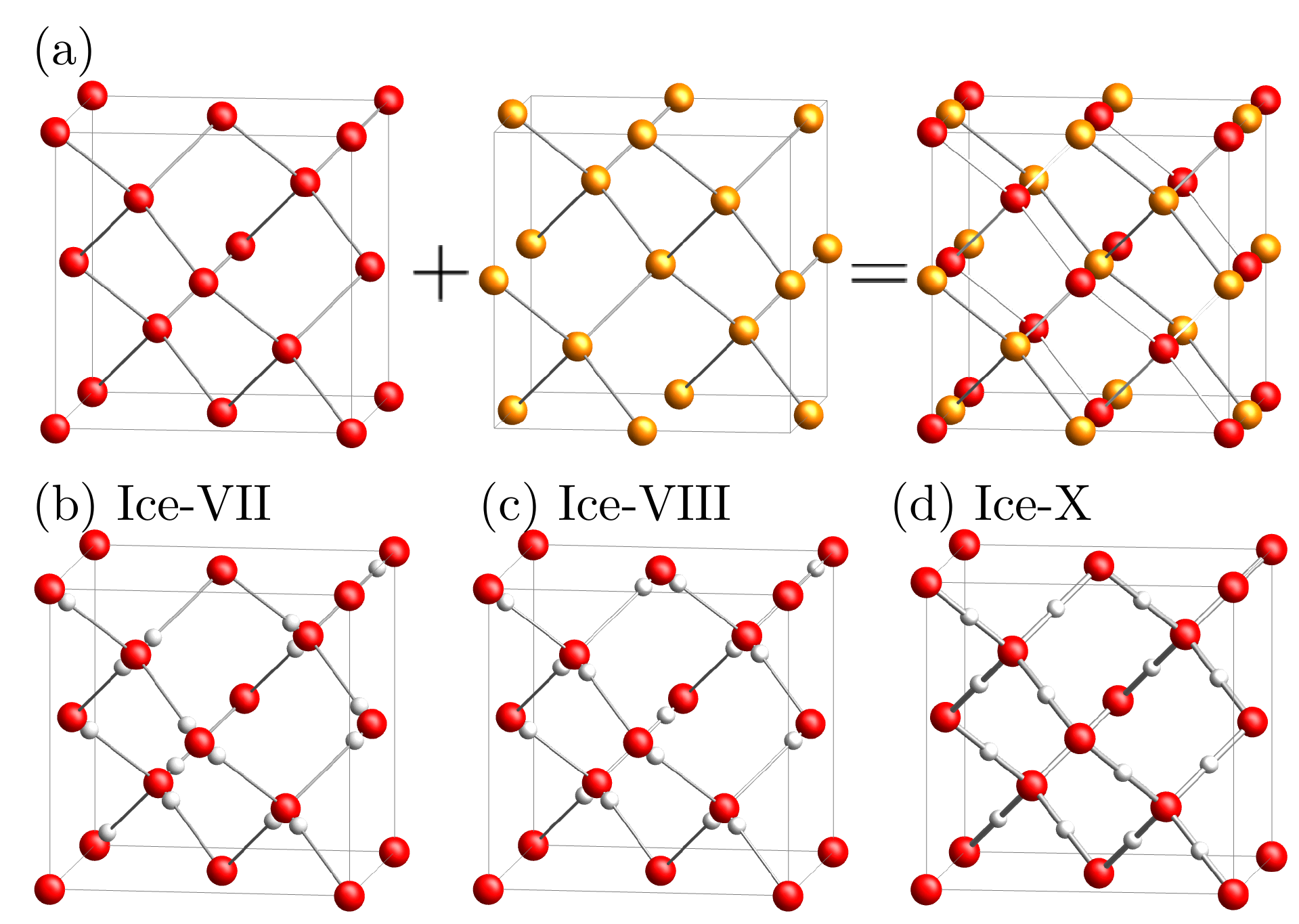}
  \caption{(a) Interpenetrating diamond structure.
  (b) The hydrogen configurations restricted to a single diamond lattice in the ice-VII phase. 
  The white spheres correspond to hydrogen atoms and the red spheres correspond to oxygen atoms. In this phase, there is no long-range order.
  (c) An example of the hydrogen configuration in the ice-VIII phase. 
  In the configuration shown, the system possesses a macroscopic polarization along the $z$-direction and the other diamond lattice has an opposite polarization.
  (d) The ice-X phase. The hydrogen atoms occupy the symmetric midpoints between oxygen atoms.}
  \label{fig:ice_structure}
\end{figure}

The crystal structure of high-pressure ice can be described by considering two primary components: the lattice formed by oxygen atoms and the spatial ordering of hydrogen atoms~\cite{Kuhs1984, Hemley1987, Salzmann2011,Komatsu02102022}.
In the ice-VII, VIII, and X phases, the oxygen atoms form an interpenetrating bipartite diamond lattice (the NaTl structure). 
Although this corresponds to a body-centered cubic (bcc) lattice in terms of the lattice points, each oxygen atom is linked to four of its eight nearest neighbors via hydrogen bonds. This network is illustrated in Fig.~\ref{fig:ice_structure}(a).
The structural parameters of high-pressure ice for this fully interpenetrating structure (space group $Pn\bar{3}m$) are summarized in Table~\ref{tab:interpenetrated}.

\begin{table*}[t]
\centering
\caption{Structural parameters for the interpenetrating structure ($Pn\bar{3}m$). Occ.\ stands for the fractional site occupancy.}
\label{tab:interpenetrated}
\renewcommand{\arraystretch}{1.5}
\begin{tabular}{lccccc}
\toprule
\multicolumn{1}{c}{Phase} & Atom & Wyckoff & Coordinates & Site Symm. & Remarks \\
\midrule
\textbf{Common to VII, X} & O & $2a$ & $(0, 0, 0)$, $(\frac{1}{2},\frac{1}{2}, \frac{1}{2})$ & $\bar{4}3m$ $(T_d)$ & bcc lattice \\
\midrule
\textbf{VII} & H & $8e$ & $(x,x,x)$, $ (\bar{x},\bar{x},x)$, & $.3m$ $(C_{3v})$ & Proton disordered \\
&&& $(\bar{x},x,\bar{x})$, $(x,\bar{x},\bar{x})$ & & (Occ. = 0.5) \\
&&& and $(x+\frac{1}{2},x+\frac{1}{2},\bar{x}+\frac{1}{2})$, etc. & & \\
\midrule
\textbf{X} & H & $4b$ & $(\frac{1}{4},\frac{1}{4},\frac{1}{4}), (\frac{3}{4},\frac{3}{4},\frac{1}{4}),$ & $\bar{3}m$ $(D_{3d})$ & Proton ordered \\
&&& $(\frac{3}{4},\frac{1}{4},\frac{3}{4}), (\frac{1}{4},\frac{3}{4},\frac{3}{4})$ & & (Centrosymmetric) \\
\bottomrule
\end{tabular}
\end{table*}

The hydrogen atoms in the lower-pressure phases (ice-VII and ice-VIII) generally obey the Bernal-Fowler ice rules~\cite{Bernal1933}:
\begin{enumerate}
  \item Each oxygen atom is coordinated with four nearest-neighbor oxygen atoms, and there is exactly one hydrogen atom on each bond connecting them.
  \item The hydrogen atom occupies an asymmetric position between two oxygen atoms, such that each oxygen atom has exactly two nearest-neighbor hydrogen atoms.
\end{enumerate}
These rules do not uniquely determine the macroscopic configuration, allowing for a macroscopically degenerate ground-state manifold characterized by the Pauling entropy~\cite{Pauling1935}. 
In ice-VII, protons are dynamically disordered within this manifold, which is crystallographically represented by the fractional occupancies of the $8e$ Wyckoff positions (Table~\ref{tab:interpenetrated}). 
In ice-VIII, a specific configuration with opposite macroscopic polarizations in each of the two interpenetrating diamond lattices is selected, resulting in long-range antiferroelectric order 
(which maps to a ferromagnetic order in our single-network effective spin model). This spontaneous polarization breaks the cubic symmetry, lowering the macroscopic space group to $I4_1/amd$.
Although our effective model is defined on a fixed, undistorted lattice, this symmetry breaking is itself correctly captured: the spontaneous ordering of the polarization (the ferromagnetic order parameter) along one of the three equivalent cubic axes breaks the $x$-$y$-$z$ equivalence. What the model does not describe is the quantitative magnitude of the accompanying lattice distortion and the somewhat higher density of ice-VIII relative to ice-VII~\cite{Kuhs1984}, which are secondary structural responses---analogous to magnetostriction---to the proton ordering rather than its driving force. Since our focus is the thermodynamic character of the proton order-disorder transformations, this simplification is not expected to affect our conclusions.
In ice-X, the ice rules are effectively broken; the hydrogen atoms are positioned at the symmetric midpoints corresponding to the $4b$ Wyckoff positions (Table~\ref{tab:interpenetrated}).

To clarify the connectivity of the hydrogen network and its underlying topology, which would otherwise be obscured by the complex interpenetrating bipartite structure, we restrict our theoretical modeling to a single, isolated diamond network. Physically, this amounts to retaining the hydrogen atoms---and the associated spin variables---of only one of the two interpenetrating networks. The two networks are crystallographically equivalent and, at the level of the hydrogen-bond topology, mutually decoupled, since every oxygen-oxygen bond belongs to a single network; modeling one network therefore faithfully captures the essential connectivity. In this single-network description, the antiferroelectric inter-network order of ice-VIII---in which the two networks carry opposite macroscopic polarizations and the bulk crystal has no net polarization---maps onto a uniform \emph{ferromagnetic} order, which is why we refer to it below as the ``ice-ferro'' state. The corresponding crystallographic parameters for this simplified single-network structure (space group $Fd\bar{3}m$) are summarized in Table~\ref{tab:single}. 
The simplified structures of ice-VII, ice-VIII, and ice-X restricted to a single network are shown in Figs.~\ref{fig:ice_structure}(b), \ref{fig:ice_structure}(c), and \ref{fig:ice_structure}(d), respectively.

\begin{table*}[t]
\centering
\caption{Structural parameters for the single-network structure ($Fd\bar{3}m$). Coordinates differing only by the face-centered cubic primitive translation vectors $(0,\tfrac{1}{2},\tfrac{1}{2})$, $(\tfrac{1}{2}, 0,\tfrac{1}{2})$, and $(\tfrac{1}{2},\tfrac{1}{2},0)$ are omitted. Occ.\ stands for the fractional site occupancy.}
\label{tab:single}
\renewcommand{\arraystretch}{1.5}
\begin{tabular}{lccccc}
\toprule
\multicolumn{1}{c}{Phase} & Atom & Wyckoff & Coordinates & Site Symm. & Corresponding structure \\
\midrule
\textbf{Common to VII, X} & O & $8a$ & $(0,0,0)$, $(\frac{3}{4}, \frac{1}{4}, \frac{3}{4})$ & $\bar{4}3m \ (T_d)$ & Diamond lattice \\
\midrule
\textbf{VII} & H & $32e$ & $(x,x,x)$, $(\bar{x}, \bar{x}+\tfrac{1}{2}, x+\tfrac{1}{2})$, & $.3m \ (C_{3v})$ & Distorted pyrochlore \\
&&& $(\bar{x}+\tfrac{1}{2}, x+\tfrac{1}{2}, \bar{x})$, $(x+\tfrac{1}{2}, \bar{x}, \bar{x}+\tfrac{1}{2})$, &  &  \\
 & & & $(x+\frac{3}{4}, x+\frac{1}{4}, \bar{x}+\frac{3}{4})$, $(\bar{x}+\frac{1}{4}, \bar{x}+\frac{1}{4}, \bar{x}+\frac{1}{4})$, & & \\
&&& $(x+\frac{1}{4}, \bar{x}+\frac{3}{4}, x+\frac{3}{4})$, $(\bar{x}+\frac{3}{4}, x+\frac{3}{4}, x+\frac{1}{4})$ & & (Occ. = 0.5, $x \approx \tfrac{1}{8}$) \\
\midrule
\textbf{X} & H & $16c$ & $(\frac{1}{8}, \frac{1}{8}, \frac{1}{8})$, $(\frac{7}{8}, \frac{3}{8}, \frac{5}{8})$, $(\frac{3}{8}, \frac{5}{8}, \frac{7}{8})$, $(\frac{5}{8}, \frac{7}{8}, \frac{3}{8})$& $\bar{3}m \ (D_{3d})$ & Ideal pyrochlore lattice \\
\bottomrule
\end{tabular}
\end{table*}

\subsection{Phase transition in dense ice}
The thermodynamic nature of the transformation from molecular ice-VII (and its low-temperature proton-ordered analog, ice-VIII) to the non-molecular, symmetric ice-X has been a central question in high-pressure physics. A prerequisite for understanding this transformation is that the temporally and spatially averaged structure of ice-VII shares the same macroscopic body-centered cubic (bcc) oxygen sublattice and $Pn\bar{3}m$ symmetry as idealized ice-X~\cite{Hemley1987}. 
Because no macroscopic symmetry is broken, an adiabatic continuous crossover between them is theoretically permissible, although a first-order phase transition is also allowed.

Experimentally, the transformation depends on the thermodynamic pathway.
At low temperatures, the compression of proton-ordered ice-VIII results in a first-order phase transition to ice-X, as evidenced by decompression hysteresis~\cite{Hirsch1986}. 
At room temperature, however, diverse techniques---including vibrational spectroscopy~\cite{Goncharov1996, Aoki1996, Struzhkin1997, Pruzan1997, Goncharov1999}, high-pressure nuclear magnetic resonance (NMR)~\cite{Meier2018}, and \textit{in situ} neutron diffraction~\cite{Guthrie2013, Komatsu2024,Komatsu02102022}---suggest that the transformation from ice-VII to ice-X occurs through a broad, continuous transition regime dominated by quantum proton tunneling and dynamic disorder.

Theoretically, various computational approaches---ranging from early lattice models~\cite{Schweizer1984} to \textit{ab initio} path-integral molecular dynamics~\cite{Bernasconi1998, Benoit1998, Benoit2002, Marques2009, Bronstein2014, Cherubini2024}---have demonstrated that nuclear quantum effects wash out classical singularities, smoothing the transformation into a continuous crossover at finite temperatures. Furthermore, a phenomenological thermodynamic model reconciled these observations by proposing that the theoretical first-order boundary between ice-VII and ice-X terminates at a critical point hidden within the low-temperature stability field of ice-VIII~\cite{Ackland2025}. Consequently, room-temperature compression bypasses this singularity, resulting in a continuous crossover characterized by ``Widom lines''---ridges of extrema in thermodynamic response functions~\cite{Mendez2021}.

\subsection{Topological phase transitions without symmetry breaking}
Given that ice-VII and ice-X share identical macroscopic symmetry, a finite-temperature phase transition between them could only occur through a non-local, topological mechanism. While rare, such transitions exist when specific conditions regarding the dimensionality and the nature of topological defects are met.

For instance, the Berezinskii-Kosterlitz-Thouless (BKT) transition in the 2D classical $XY$ model is driven by the topological unbinding of vortex pairs, leading to an essential singularity in the correlation length~\cite{Kosterlitz1974}. This occurs without spontaneous symmetry breaking, as such breaking is prohibited by the Mermin-Wagner theorem~\cite{Mermin1966, Berezinskii1971, Kosterlitz1973}. In 3D systems, such as the Kitaev model~\cite{Kitaev2003, Kitaev2006, PhysRevB.90.205126, Nasu2014}, the 3D toric code~\cite{Castelnovo2008_Toric}, or its variants with explicitly broken symmetries~\cite{Polyakov1978, Fradkin1979, Hastings2011}, macroscopic phase transitions can survive at finite temperatures. This survival occurs because the relevant topological defects take the form of closed one-dimensional loops~\cite{Wegner1971}. Unlike point defects, these extended 1D loops can deform around local 0D thermal point defects in 3D space without losing their topological information~\cite{Castelnovo2008_Toric,PhysRevB.90.205126}. 
Their line tension suppresses their thermal generation, thereby protecting a classical topological order up to a critical temperature.  Furthermore, a recent study has shown that macroscopic phase transitions can survive even with point-like defects if they obey anomalous fermionic statistics, as demonstrated in the 3D fermionic toric code~\cite{Zhou2025}.

In contrast, the topological defects in high-pressure ice---namely, violations of the ice rules (ionic defects) acting as effective magnetic monopoles~\cite{Castelnovo2008, Benton2016}---are classical point-like excitations in 3D real space, lacking such geometric or quantum statistical protection. While the effective spin model investigated in this study is classical and does not incorporate nuclear quantum effects---which can theoretically stabilize a quantum spin ice state at absolute zero~\cite{Benton2016}---the topological fragility at finite temperatures is applicable. In standard 3D spin ice models, whether classical or quantum, point-like bosonic monopoles proliferate at finite temperatures, forming a magnetic plasma. This induces Debye-H\"{u}ckel screening that destroys the long-range dipolar correlations (and any associated macroscopic entanglement in quantum analogs) characteristic of the Coulomb phase~\cite{Polyakov1977, Henley2005, Castelnovo2011, Otsuka2014, Benton2016}. As established in the 3D classical $\mathbb{Z}_2$ gauge-Higgs theory~\cite{Fradkin1979}, the presence of unconfined dynamic point-like defects acts analogously to a fluctuating matter (Higgs) field, bridging the confined and deconfined regimes via Higgs-confinement continuity. This vulnerability of standard point-like defects to thermal fluctuations reinforces the theoretical expectation that the transformation between ice-VII and ice-X is a continuous crossover rather than a phase transition.

\section{Effective Spin Model}
\label{sec:model}

\subsection{Ice model}
\begin{figure}[t]
  \centering
  \includegraphics[width=\linewidth]{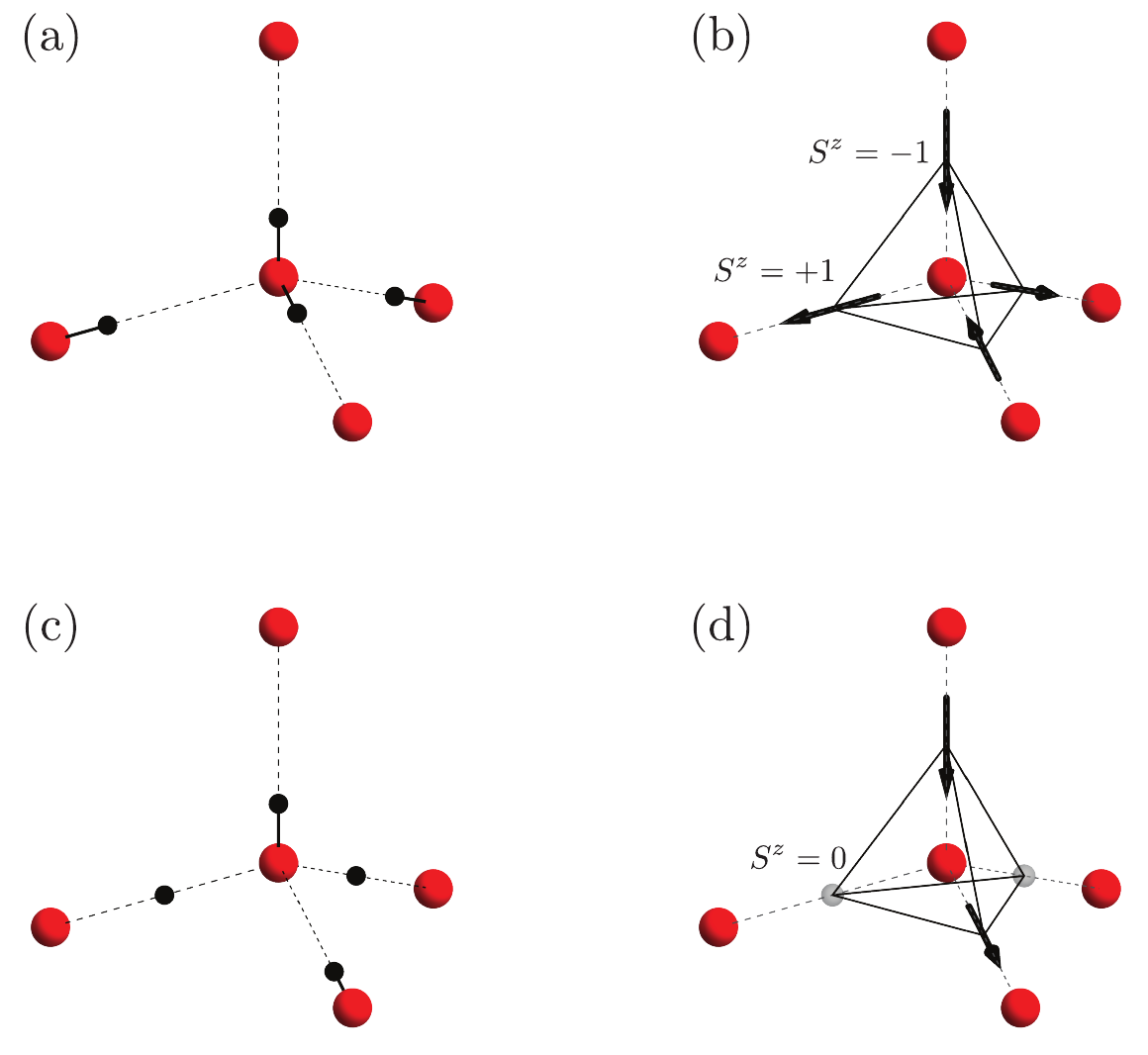}
  \caption{Mapping between the proton positions and the effective spin variable, following the standard spin-ice convention~\cite{Harris1997, Bramwell2001} and extended to the spin-$1$ case relevant here. (a) A ``two-in, two-out'' proton configuration; the large spheres are oxygen atoms and the black dots are protons, each displaced off-center along an oxygen-oxygen bond. (b) The corresponding spin configuration on a tetrahedron of pyrochlore sites. With the local quantization axis taken along each bond from the $A$ to the $B$ diamond sublattice, a proton displaced toward the central ($A$-sublattice) oxygen corresponds to $S^z=-1$ and one displaced toward the outer ($B$-sublattice) oxygen to $S^z=+1$. (c) A configuration in which two of the protons instead sit at the symmetric bond midpoints, as in ice-X. (d) The corresponding spins, in which the centered protons map to the $S^z=0$ state while the off-center protons retain $S^z=\pm1$.}
  \label{fig:mapping}
\end{figure}

\begin{figure}[t]
  \centering
  \includegraphics[width=\linewidth]{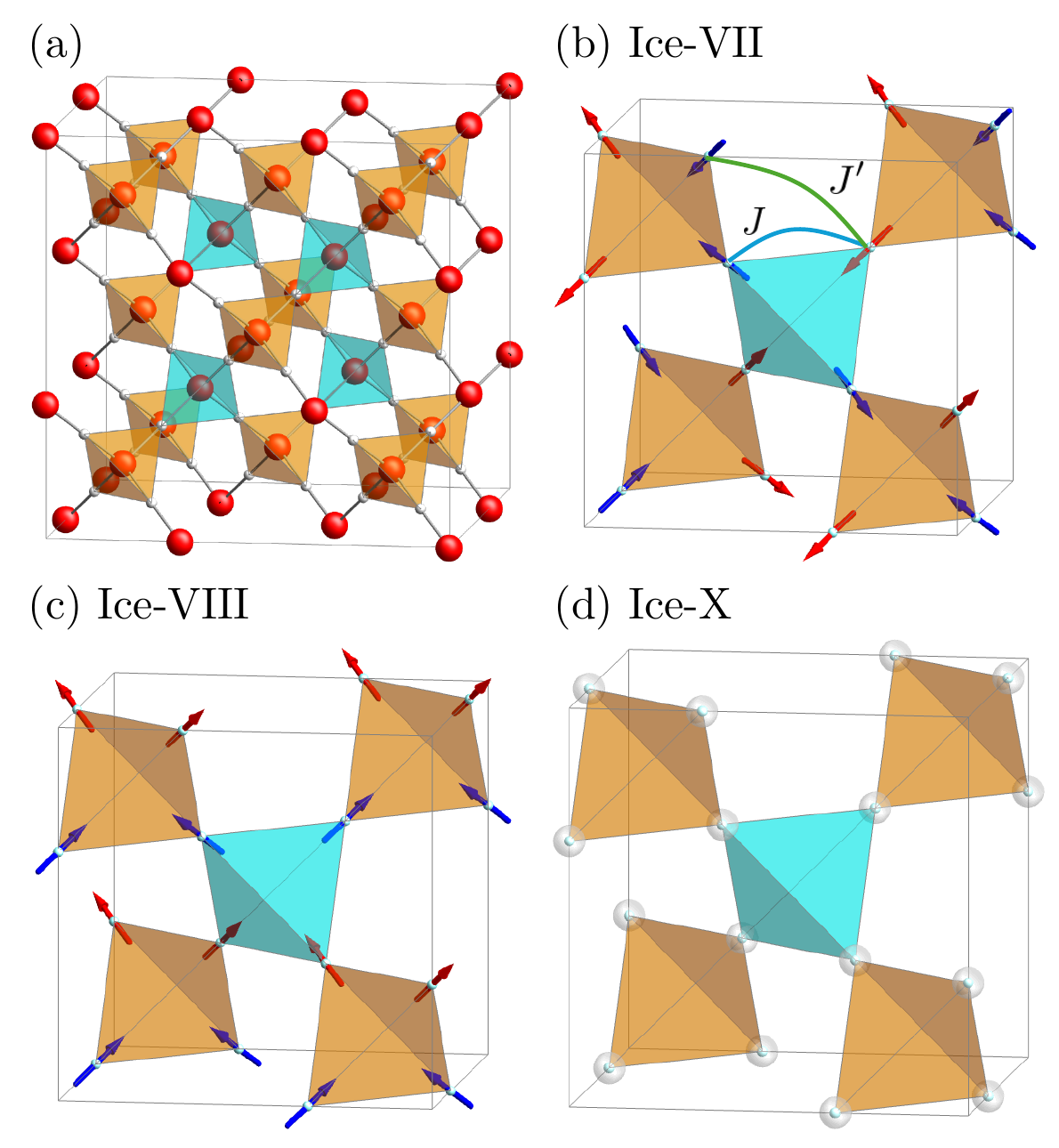}
  \caption{(a) An illustration of the relationship between the diamond lattice and the pyrochlore lattice. The two colors of tetrahedra correspond to the $A$ and $B$ sublattices of the underlying diamond lattice (the ``up'' and ``down'' tetrahedra of the pyrochlore lattice).
  (b) The spin representation of the hydrogen configuration in the ice-VII phase. Each tetrahedron satisfies the ``two-in, two-out'' configuration but may possess a different local magnetic moment. The nearest-neighbor coupling $J$ and the next-nearest-neighbor coupling $J'$ are indicated by the arcs.
  (c) The ice-VIII phase. Each tetrahedron satisfies the ice rules and shares an identical macroscopic magnetic moment aligned along the $z$-direction.
(d) The ice-X phase. The hydrogen atoms are localized at the symmetric midpoints of the oxygen-oxygen bonds, which corresponds to the $S^z=0$ state on all pyrochlore sites. 
  }
  \label{fig:pyrochlore_structure}
\end{figure}

To construct a tractable statistical mechanics model, we map the proton configurations onto an effective spin model defined on the pyrochlore lattice.
Since the ice rules require hydrogen atoms to occupy asymmetric positions along the oxygen-oxygen bonds in the lower-pressure phases, their two possible localized positions can be represented by an Ising variable $\sigma^z_\ell \in \{-1, 1\}$ assigned to each pyrochlore site $\ell$. This correspondence between the proton positions and the spin states is illustrated in Figs.~\ref{fig:mapping}(a) and \ref{fig:mapping}(b). 
The symmetric midpoints of the hydrogen bonds form a pyrochlore lattice, as illustrated in Fig.~\ref{fig:pyrochlore_structure}(a).
As detailed in Table~\ref{tab:single}, these symmetric midpoints correspond to the hydrogen positions in the idealized ice-X phase (the $16c$ Wyckoff positions), which natively constitute an ideal pyrochlore lattice.

The conventional unit cell of this underlying lattice is cubic and contains 16 sites.
Assuming the cubic lattice constant is unity ($a=1$), the full lattice is generated by translating the four basis sites corresponding to these $16c$ positions---$(\frac{1}{8}, \frac{1}{8}, \frac{1}{8})$, $(\frac{7}{8}, \frac{3}{8}, \frac{5}{8})$, $(\frac{3}{8}, \frac{5}{8}, \frac{7}{8})$, and $(\frac{5}{8}, \frac{7}{8}, \frac{3}{8})$---by the face-centered cubic primitive translation vectors:
\begin{align}
  \bm{a}_1 &= (0, 1/2, 1/2),\\
  \bm{a}_2 &= (1/2, 0, 1/2),\\
  \bm{a}_3 &= (1/2, 1/2, 0).
\end{align}
This geometry yields a total of $N=16L^3$ sites for a system with linear dimension $L$.
The four basis sites form a regular tetrahedron.

In the extended lattice, the sites form a network of corner-sharing tetrahedra that can be bipartitioned into two alternating sets (often referred to as ``up'' and ``down'' tetrahedra).
The centers of one set of tetrahedra correspond to the A sublattice of the underlying diamond lattice, which consists of the lattice points generated by translations of the $(0,0,0)$ position as detailed in Table~\ref{tab:single}. The centers of the other set correspond to the B sublattice, consisting of the lattice points generated by translations of the $(\frac{3}{4}, \frac{1}{4}, \frac{3}{4})$ position.  
Here, we explicitly define the local quantization axis for the spin variables. We align the local $z$-axis along each oxygen-oxygen bond such that it points from the A sublattice to the B sublattice. In this convention, a spin-up state ($\sigma_\ell^z = +1$) indicates that the proton is physically displaced toward the oxygen atom on the B sublattice, whereas a spin-down state ($\sigma_\ell^z = -1$) corresponds to the proton being displaced toward the A sublattice. Each of the four sites of a tetrahedron carries its own local quantization axis pointing along one of the four $\langle 111 \rangle$ cubic directions (the oxygen-oxygen bond directions); thus $\sigma_\ell^z=\pm1$ is defined relative to these site-dependent local axes rather than a single global axis. Since all exchange couplings are identical on symmetry-related bonds and the single-ion anisotropy introduced below is uniform, the Hamiltonian retains the full cubic point-group symmetry of the lattice, including the four-fold ($90^\circ$) rotations about the $\langle 100 \rangle$ axes. With this physical correspondence, the Bernal-Fowler ice rules are equivalent to the ``two-in, two-out'' spin configurations on each tetrahedron [Fig.~\ref{fig:pyrochlore_structure}(b)], realized as the degenerate ground states of the nearest-neighbor antiferromagnetic Ising model~\cite{Harris1997, Bramwell2001}:
\begin{align}
  H = J\sum_{\langle \ell,\ell' \rangle}\sigma_\ell^z\sigma_{\ell'}^z \quad (J > 0).
\end{align}
Here, $\langle \ell,\ell' \rangle$ denotes summation over nearest-neighbor spin pairs, i.e., the six edges within each tetrahedron of the pyrochlore lattice.

In real water ice crystals, the macroscopic degeneracy of the hydrogen-bond network is dynamically explored through the creation and migration of two types of thermally excited point defects: Bjerrum defects (pairs of an empty hydrogen bond, known as a Leere or L-defect, and a doubly occupied bond, known as a Doppel or D-defect) and ionic defects (pairs of $\mathrm{H_3O^+}$ and $\mathrm{OH^-}$ ions). These defects behave as mobile topological charges migrating through the ice~\cite{Ryzhkin1983}. Through their diffusion, the system transitions between configurations that locally satisfy the ``two-in, two-out'' ice rules, resulting in the dynamic disorder characteristic of ice-VII~\cite{Bernal1933, Pauling1935}. In the language of spin models, this proton disordered state lacks long-range order, serving as a classical structural analog to ice-VII, and exhibits topological quasi-long-range order characterized by algebraic dipolar correlations and pinch-point singularities indicative of a Coulomb phase~\cite{Huse2003, Isakov2004, Benton2016}.

To realize the long-range ordered phase (ice-VIII), further interactions lifting the macroscopic degeneracy of the ice rules must be introduced. In theoretical models of hydrogen-bonded ferroelectrics, macroscopic polarization is stabilized by breaking the emergent gauge symmetry via longer-range interactions, such as dipole-dipole couplings~\cite{Chern2014}. 
Following this mechanism, we introduce a positive next-nearest-neighbor interaction $J'$ to stabilize the macroscopic polarization corresponding to the ice-VIII phase (sometimes referred to as the ``ice-ferro'' state~\cite{Ishizuka2014}):
\begin{align}
  H = J\sum_{\langle \ell,\ell' \rangle}\sigma_\ell^z\sigma_{\ell'}^z + J'\sum_{\{\ell,\ell'\}}\sigma_\ell^z\sigma_{\ell'}^z.
\end{align}
Here, $\{\ell,\ell'\}$ denotes the next-nearest-neighbor spin pairs of the pyrochlore lattice---sites that do not share a tetrahedron but lie on two tetrahedra joined at a common corner; the $J$ and $J'$ bonds are illustrated in Fig.~\ref{fig:pyrochlore_structure}(b). A positive $J'$ lifts the macroscopic ice-rule degeneracy and selects, out of the ``two-in, two-out'' manifold, the uniformly polarized ($\bm{q}=0$) ice-ferro configuration.
The spin configuration of the ice-ferro state is presented in Fig.~\ref{fig:pyrochlore_structure}(c).
This configuration possesses a macroscopic magnetic moment, which corresponds to the macroscopic electric polarization of the hydrogen-bond network in ice-VIII.

We emphasize that the $J'$ term is a minimal, effective device for lifting the ice-rule degeneracy in favor of the polarized ($\bm{q}=0$) state. Physically, the antiferroelectric order of bulk ice-VIII is stabilized by the (predominantly electrostatic) coupling between the two interpenetrating networks, which carry opposite polarizations; $J'$ encodes the net thermodynamic effect of this inter-network coupling---the selection of the uniformly polarized single-network configuration---rather than asserting a literal next-nearest-neighbor interaction within a single network. Our aim is not to reproduce the microscopic origin of this coupling, but to retain its essential consequence, namely the first-order character of the ice-VIII boundary.

\subsection{Blume-Capel model on the pyrochlore lattice}
To capture the transition to ice-X where protons sit at symmetric midpoints [Figs.~\ref{fig:mapping}(c) and \ref{fig:mapping}(d)], we expand our local Hilbert space to a spin-$1$ model on the pyrochlore lattice, mapping the symmetric position to a state with $S^z_\ell = 0$~\cite{Pandey2025}:
\begin{align}
  H = J\sum_{\langle \ell,\ell' \rangle} S_\ell^z S_{\ell'}^z + J'\sum_{\{\ell,\ell'\}} S_\ell^z S_{\ell'}^z + \Delta \sum_\ell (S_\ell^z)^2, \label{eq:BC}
\end{align}
where $\Delta$ acts as a chemical potential penalizing asymmetric proton states. A positive $\Delta$ drives the system into the proton-centered ice-X phase.

Because our effective spin formulation assigns one spin variable ($S^z_\ell$ or $\sigma^z_\ell$) to each hydrogen bond, it imposes the condition that every oxygen-oxygen bond accommodates exactly one proton. Consequently, this model operates in a limit where the formation of Bjerrum defects is prohibited. On the other hand, local violations of the ice rules on the tetrahedra---such as ``3-in, 1-out'' or ``1-in, 3-out'' spin configurations---correspond to the formation of $\mathrm{H_3O^+}$ and $\mathrm{OH^-}$ ionic defects, respectively. These ionic defects act as unconfined topological point defects, treated as magnetic monopoles within the spin-ice framework~\cite{Castelnovo2008, Castelnovo2011, Benton2016}. The energy cost of creating such an ionic defect is set by the nearest-neighbor coupling $J$: relative to a ``two-in, two-out'' tetrahedron, a ``three-in, one-out'' (or ``one-in, three-out'') tetrahedron costs an energy $2J$, so that flipping a single spin in an ice-rule background creates a defect pair at a cost of $4J$. Ionic defects are thus thermally activated rather than free, with a density controlled by $J/T$; they are absent only in the strict $T/J\to0$ limit.

In the high-pressure regime approaching the transition to the ice-X phase, the proton dynamics are predominantly governed by intra-bond displacement rather than inter-bond rotational hopping, consistent with the hydrogen-bond symmetrization observed under pressure~\cite{Goncharov1996, Benoit1998}. Thus, our theoretical approach of capturing ionic defects while projecting out Bjerrum defects is physically motivated.

A physically important process in dense ice is, in particular, the empty-bond (L) defect, in which a proton vacates an intra-network bond and reorients toward the neighboring \emph{interpenetrating} network; such inter-network excursions intrinsically couple the two diamond networks and therefore lie outside the present single-network description. This omission is justified in the regime of interest. The two networks are crystallographically independent and are not hydrogen-bonded to one another~\cite{Kuhs1984, Salzmann2011}, and inter-network proton configurations become statistically significant only in the high-temperature ``plastic'' regime near the melting line, where rapid molecular reorientation produces pronounced sublattice mixing~\cite{Rescigno2025}. In the proton-disordered, proton-ordered, and symmetric solid phases (ice-VII, VIII, and X) that are the subject of this study, such inter-network configurations remain rare excitations, so that the single-network model is expected to capture the essential thermodynamics faithfully.

As discussed in Sec.~\ref{sec:intro}, theoretical studies of the pure nearest-neighbor version ($J'=0$) of this model in the monopole-free limit ($T/J \to 0$) have identified topological phase transitions separating three distinct regimes: a paramagnetic phase and two topologically distinct Coulomb phases~\cite{Pandey2025}. In the context of high-pressure ice, the paramagnetic phase and the $\mathbb{Z}_2$ flux-confined Coulomb phase theoretically correspond to the idealized ice-X and ice-VII phases, respectively.

A recent study~\cite{Watanabe2026} investigated the finite-temperature fate of these topological phases in the pure nearest-neighbor model ($J'=0$). 
Through duality mappings to 3D $XY$ and Ising loop-gas models, it was demonstrated that the topological phase boundaries present in the monopole-free limit degrade into continuous crossovers at any finite temperature. Specifically, thermal monopoles act as a symmetry-breaking field in the continuous $XY$ picture and topologically sever defect strings in the Ising loop-gas picture. 
Building on these analytical insights, here we study the finite-temperature phase diagram among ice-VII, VIII, and X using Monte Carlo methods based on Eq.~\eqref{eq:BC}. This incorporates both the thermal monopole excitations inevitable in real ice crystals at finite temperatures and the next-nearest-neighbor interaction $J'$ necessary to stabilize the ice-VIII phase.

\section{Method}
\label{sec:methods}
To distinguish the different phases and crossover regimes in our effective spin model, we define appropriate order parameters and structural indicators.
To characterize the macroscopic symmetry breaking corresponding to ice-VIII, we define a ferromagnetic order parameter $\bm{M}$ (which corresponds to the macroscopic polarization vector in the ice):
\begin{equation}
    \bm{M} \coloneqq \frac{1}{4 L^3}\sum_{\ell = \langle \bm{r}_A, \bm{r}_B \rangle} (\bm{r}_B  - \bm{r}_A)S^z_\ell,
\end{equation}
where $\bm{r}_A$ and $\bm{r}_B$ denote the centers of the adjacent tetrahedra (belonging to the A and B sublattices of the diamond lattice, respectively) connected by the link $\ell$. Its long-range order 
\begin{equation}
m_{\mathrm{VIII}}\coloneqq \expval{\bm{M}^2}
\end{equation}
becomes nonzero whenever the ferromagnetic order develops and saturates at $m_{\mathrm{VIII}}=1$ in the fully polarized state.

The $S^z=0$ occupation serves as a structural indicator for the ice-X phase:
\begin{align}
 m_\mathrm{X} \coloneqq 1-\frac{1}{N}\sum_\ell \expval{(S_\ell^z)^2}.
\end{align} 

To identify phase boundaries and crossovers toward ice-X, we calculate the specific heat $c$:
\begin{align}
  c \coloneqq \frac{1}{N}\frac{\expval{E^2}-\expval{E}^2}{T^2},
\end{align}
alongside the susceptibility $\chi_\mathrm{X}$ associated with $m_\mathrm{X}$:
\begin{align}
  \chi_\mathrm{X} \coloneqq \frac{1}{NT} \Big( \Big\langle\Big(\sum_\ell (S_\ell^z)^2\Big)^2\Big\rangle- \Big\langle\sum_\ell (S_\ell^z)^2\Big\rangle^2 \Big).
\end{align}

We investigate the thermodynamic phase diagram via classical Monte Carlo (MC) simulations of the model in Eq.~\eqref{eq:BC}. 
The total number of sites is $N=16L^3$, where $L$ is the linear system size. 
One MC step consists of $N$ single-spin updates using the heat-bath method. 
Thermalization is performed for $10^6$ MC steps, followed by measurements accumulated over $5\times 10^6$ MC steps. 
For small $\Delta$, we utilize a hybrid method combining the short-loop algorithm~\cite{Barkema1998,Melko2001} and the heat-bath method to accelerate relaxation. 
Near the first-order phase transition, we employ replica exchange Monte Carlo with respect to $\Delta$ to enhance sampling. Additionally, to mitigate hysteresis and precisely locate the phase boundary, we prepare mixed initial configurations in which half of the system is initialized in an ordered state and the other half in a disordered state.

\section{Results}
\label{sec:results}

\subsection{Transition between ice-VII and ice-X}

To investigate the transition between ice-VII and ice-X, we performed numerical calculations with $J'=0$ for various temperatures $T$ and penalty parameters $\Delta$, fixing $J =1$.

\begin{figure}[t]
  \centering
  \includegraphics[width=\linewidth]{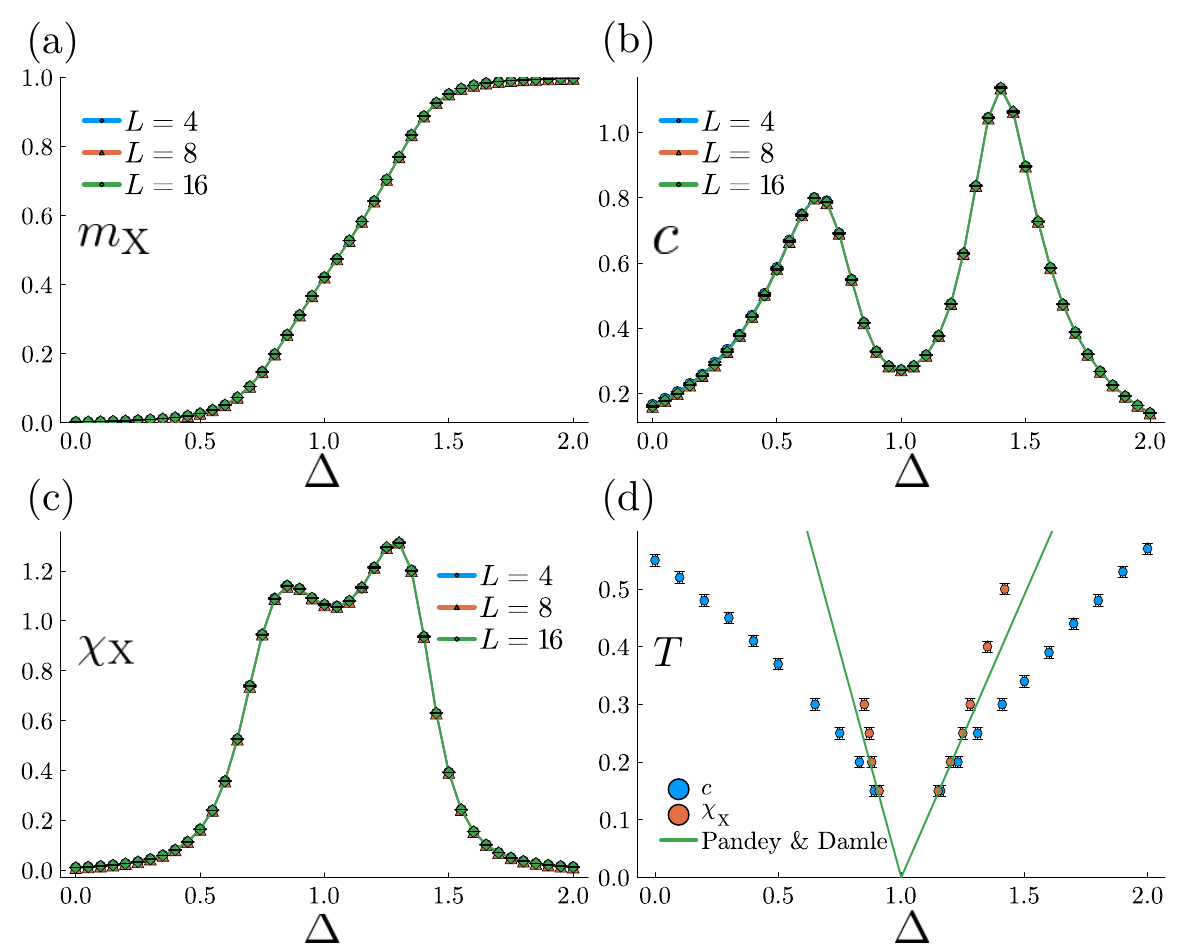}
  \caption{Monte Carlo results for the transformation between ice-VII and ice-X ($J'=0$).
  (a) The $\Delta$ dependence of the $S^z=0$ occupation $m_\mathrm{X}$ at $T=0.3$ and $L=16$.
  (b) The system size dependence of the specific heat $c$ and (c) the susceptibility $\chi_\mathrm{X}$ at $T=0.3$. The peak heights saturate as $L$ increases, indicating a continuous crossover rather than a phase transition. Distinct symbols denote $L=4$, $8$, and $16$, whose data coincide almost perfectly.
  (d) The peak positions of the second-order derivatives of the free energy for $L=4$. 
  Blue points show peaks of $c$, while orange points represent peaks of $\chi_\mathrm{X}$. 
  The green lines indicate the phase transition boundaries in the monopole-free limit~\cite{Pandey2025}.}
  \label{fig:vii_x}
\end{figure}

The $\Delta$ dependence of the $S^z=0$ occupation $m_\mathrm{X}$ at $T=0.3$ is shown in Fig.~\ref{fig:vii_x}(a). 
As $\Delta$ increases, the $S^z=0$ occupation $m_\mathrm{X}$ approaches unity, indicating a gradual crossover to the ice-X phase. 
We checked whether this structural change constitutes a thermodynamic phase transition by analyzing the second derivatives of the free energy. 
The specific heat $c$ and the susceptibility $\chi_\mathrm{X}$ are plotted as functions of $\Delta$ for different system sizes $L$ in Figs.~\ref{fig:vii_x}(b) and \ref{fig:vii_x}(c). 
For a continuous phase transition, thermodynamic response functions should diverge in the thermodynamic limit. 
However, our results demonstrate that the peaks of $\chi_\mathrm{X}$ and specific heat $c$ do not diverge but saturate to finite values. Indeed, the data for $L=4$, $8$, and $16$ in Figs.~\ref{fig:vii_x}(a)--\ref{fig:vii_x}(c) overlap almost perfectly; this absence of any appreciable system-size dependence is itself the hallmark of a crossover rather than a phase transition.

Furthermore, the peak positions of the specific heat $c$ and the susceptibility $\chi_\mathrm{X}$ do not coincide at finite temperatures [Fig.~\ref{fig:vii_x}(d)], and they clearly deviate from the phase transition boundaries predicted in the monopole-free limit~\cite{Pandey2025}.
While peak positions for different thermodynamic quantities can differ at finite system sizes even near a genuine continuous phase transition, they must systematically converge to a single critical point in the thermodynamic limit. In our results, the peaks remain separated without showing any systematic tendency to converge, which, together with the lack of divergence, signifies a continuous crossover.

These observations suggest that there is no thermodynamic phase transition separating ice-VII and ice-X for $T \gtrsim 0.1$. Instead, the system undergoes a continuous crossover. 
At absolute zero ($T=0$), a first-order phase transition point appears at $\Delta=J$ due to an energy level crossing. We emphasize that, since $J'=0$ here, no ice-VIII (ferro) order is selected; the relevant competition at $T=0$ is between the macroscopically degenerate ``two-in, two-out'' manifold corresponding to ice-VII and the proton-symmetric ice-X state. With every spin satisfying $|S^z_\ell|=1$, any ice-rule configuration has energy $E_\mathrm{VII}=(\Delta-J)N$, whereas the fully symmetric ice-X state ($S^z_\ell=0$ on all sites) has $E_\mathrm{X}=0$. These cross at $\Delta=J$, where the entire ice-VII manifold becomes degenerate with ice-X, which is the origin of the $T=0$ first-order point. However, this strict singularity does not extend to finite temperatures; the introduction of thermal fluctuations immediately smooths the transformation into a continuous crossover at any $T>0$.

\begin{figure}[t]
  \centering
  \includegraphics[width=\columnwidth]{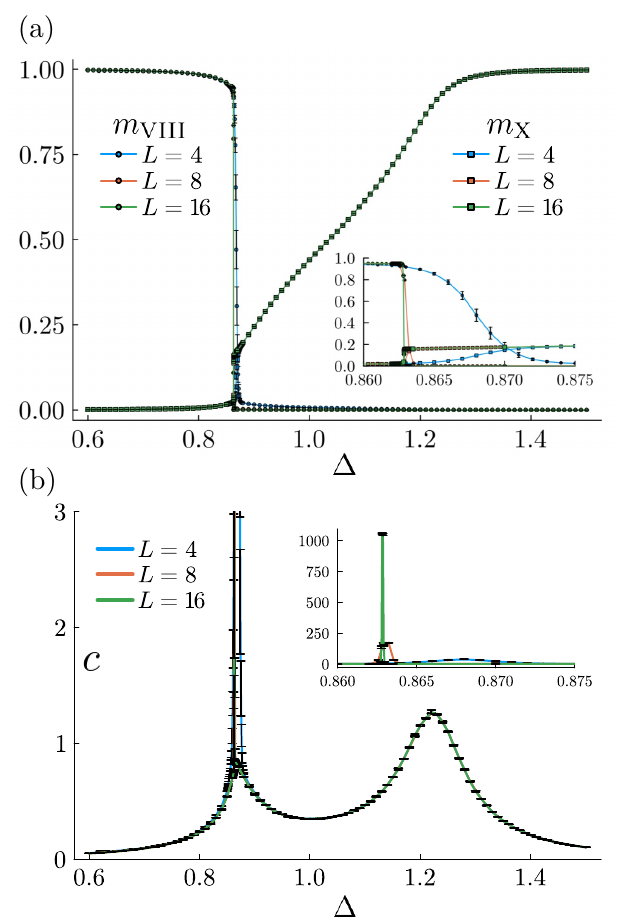}
\caption{
Monte Carlo results for $J' = 0.02$ and $T = 0.2$. 
System size dependence of (a) the squared ferromagnetic order parameter $m_{\mathrm{VIII}}$ (circles) and the $S^z = 0$ occupation $m_{\mathrm{X}}$ (squares), and (b) the specific heat $c$. 
The insets show zoomed-in views near the phase transition point at $\Delta \simeq 0.86$. 
}
  \label{fig:viii_x}
\end{figure}

\subsection{Phase transitions among ice-VII, VIII, and X}

Next, we set $J'=0.02$ to investigate the phase transitions involving ice-VIII. The $\Delta$ dependence of the structural indicator $m_\mathrm{X}$ and the ferromagnetic order parameter $m_{\mathrm{VIII}}$ at $T=0.2$ is shown in Fig.~\ref{fig:viii_x}(a). 
The squared ferromagnetic order parameter $m_{\mathrm{VIII}}$ drops discontinuously to zero at $\Delta \simeq 0.86$, signifying a first-order phase transition destroying the long-range ice-ferro order. 
The corresponding specific heat exhibits a delta-functional peak at the transition [Fig.~\ref{fig:viii_x}(b)]. Note that this peak scales approximately with the system volume $L^3$, supporting the first-order nature of the transition destroying the ice-VIII order.

Simultaneously, a jump in $m_\mathrm{X}$ is observed, although the occupation does not immediately saturate to unity.
Once the ice-VIII order is destroyed, the system enters a disordered phase, which continuously connects both the ice-VII state at smaller $\Delta$ and the proton-symmetric ice-X phase at larger $\Delta$ without any further jumps. 
This implies that the ice-VIII state does not transform directly and continuously into ice-X; it must first undergo a first-order phase transition into the disordered phase, which subsequently undergoes a continuous crossover into ice-X. 
In contrast to the sharp transition, the broad peak in $c$ around $\Delta \approx 1.2$ remains completely independent of the system size, which clearly indicates this continuous crossover toward the ice-X state.

\begin{figure}[t]
  \centering
  \includegraphics[width=\columnwidth]{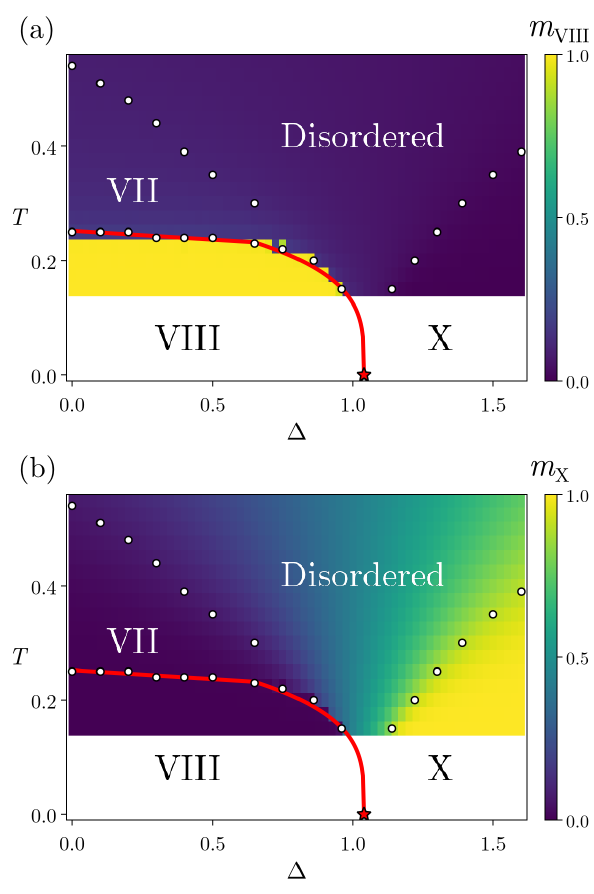}
  \caption{Phase diagram for $J'=0.02$ and $L=4$.
  The color maps represent (a) the ferromagnetic order parameter $m_{\mathrm{VIII}}$ and (b) the $S^z=0$ occupation $m_\mathrm{X}$.
  White circles denote peak positions of the specific heat $c$, and the red star represents the first-order transition point at $T=0$ arising from a level crossing. At $T=0$, the level crossing point is shifted to $\Delta_c = J + 2J' = 1.04$ due to the energy gain from the next-nearest-neighbor interaction.
  The red line, drawn as a guide to the eyes, represents the expected first-order phase transition boundary separating the ice-VIII phase from the other phases.}
  \label{fig:phase_diagram_J2_002}
\end{figure}

The global phase diagram, derived from the specific heat peaks overlaid on the color maps of $m_{\mathrm{VIII}}$ and $m_\mathrm{X}$, is plotted in Fig.~\ref{fig:phase_diagram_J2_002}. Notably, direct phase boundaries separating the ice-VII and ice-X states, or separating the ice-VIII and ice-X states, are entirely absent. Instead, a broad disordered regime universally intervenes at finite temperatures. Furthermore, unlike the liquid-gas transition discussed in Sec.~\ref{sec:intro}, within the investigated parameter space and the resolution of our Monte Carlo simulations, we observe no indication of a finite-temperature first-order phase transition line terminating at a critical point separating the ice-VII and ice-X regimes. A genuine thermodynamic phase transition at finite temperatures is therefore exclusively associated with the destruction of the ice-VIII phase.

\section{Discussion and Conclusion}
\label{sec:discussion}

Our numerical results demonstrate that the model states corresponding to ice-VII and ice-X are adiabatically connected via a continuous crossover regime with an intervening disordered phase at finite temperatures. While this broad intervening disordered regime might initially appear to contrast with the nominal classification of ice-VII and ice-X as distinct adjacent solid phases, it actually aligns well with recent room-temperature experiments. These experiments suggest a wide transitional zone dominated by dynamic proton fluctuations and quantum tunneling before the system fully settles into the completely symmetric ice-X phase~\cite{Goncharov1996, Aoki1996, Struzhkin1997, Pruzan1997, Goncharov1999, Meier2018, Guthrie2013, Komatsu2024}. Our classical effective model physically captures this intermediate regime as a topological Coulomb liquid fundamentally smeared by thermal monopoles. Because ice-VII and ice-X belong to the same macroscopic space group and lack any local order parameter distinguishing them, they become thermodynamically indistinguishable once thermal fluctuations are accounted for. We substantiate this physical picture by drawing a comparison with theoretical studies on the $S=1$ pyrochlore magnet and the general topological principles discussed in Sec.~\ref{sec:background}.

At absolute zero, theoretical studies assuming an enforcement of the ice rules (the monopole-free limit) have established the existence of topological phase transitions and exotic quantum phases, such as those found in the $S=1$ pyrochlore magnet~\cite{Pandey2025}. 
In the broader context of water ice, the mapping of the Bernal-Fowler ice rules to an emergent $U(1)$ gauge theory has demonstrated that collective quantum tunneling of protons can theoretically stabilize a $U(1)$ quantum liquid state endowed with emergent photons at low temperatures~\cite{Benton2016}. Building upon such concepts, a theoretical study by Han and Kivelson~\cite{Han2025} demonstrated that strong electron-phonon couplings in band insulators can stabilize a quantum phononic resonating-valence-bond (pRVB) phase featuring an emergent $U(1)$ gauge field, suggesting water ice under pressure as a candidate material for such zero-temperature topological orders.

However, our finite-temperature MC simulations, which explicitly include thermal monopole excitations, reveal that as soon as thermal fluctuations are introduced, these putative topological phase boundaries are replaced by continuous crossovers. Translated to the context of water ice, this implies that regardless of the topological possibilities at $T=0$, the physical system at finite temperatures exhibits a continuous transformation from ice-VII, through an intermediate disordered regime, to ice-X.

We attribute this qualitative shift to the screening effect by thermal magnetic monopoles. While the zero-temperature quantum and classical studies rely on the prohibition of monopoles, the thermal excitation of monopoles (violations of the ice rules) is thermodynamically inevitable at any finite temperature $T>0$. As highlighted in Sec.~\ref{sec:background}, the topological Coulomb phase is unstable at $T>0$ in the presence of dynamic, point-like matter fields~\cite{Polyakov1977,Henley2005,Castelnovo2011,Otsuka2014}. This is a consequence of Polyakov's no-go theorem~\cite{Polyakov1977} for three-dimensional compact gauge theories, which dictates that confinement is unavoidable at finite temperatures.

This topological fragility induced by point-like defects is demonstrated in a recent analytical study on the pure $S=1$ pyrochlore spin ice~\cite{Watanabe2026}. By mapping the system onto dual 3D $XY$ and Ising loop-gas representations, we demonstrated that thermal monopoles act as a symmetry-breaking field in the continuous $XY$ picture, and geometrically act as endpoints that sever the defect strings in the Ising loop-gas picture. Consequently, the topological phase transitions are smeared into continuous crossovers at any finite temperature, aligning with our present numerical observations.

The inevitability of this crossover highlights a distinction regarding the nature of topological defects. As discussed in Sec.~\ref{sec:background}, topological phase boundaries can be shielded from screening if the underlying defects are extended loop-like objects, such as in the Kitaev model and the bosonic 3D toric code~\cite{Castelnovo2008_Toric}, or if point-like defects possess anomalous quantum statistics, as in the fermionic toric code~\cite{Zhou2025}. In contrast, the defects in our classical ice system are point-like and lack such geometric or quantum statistical protection. It is established in spin ice models that the algebraic dipolar correlations and sharp pinch points characteristic of the Coulomb phase are smeared out and replaced by exponential decay due to Debye-H\"{u}ckel screening by the monopole plasma~\cite{Henley2005,Castelnovo2011,Otsuka2014}. Consequently, the topological phase transitions defining the $\mathbb{Z}_2$ flux-confined and deconfined phases cannot survive once the thermodynamic proliferation of point-like monopoles is taken into account.

The dichotomy between the continuous ice-VII to ice-X crossover and the first-order phase transition into the proton-ordered ice-VIII phase can be understood through the lens of explicit symmetry breaking. As demonstrated in quantum Monte Carlo studies of hydrogen-bonded ferroelectrics~\cite{Chern2014}, the introduction of longer-range interactions (analogous to our $J'$ term) explicitly breaks the macroscopic gauge symmetry of the ice rules, replacing the underlying topological transition with a symmetry-breaking phase transition. In the low-temperature regime ($0<T\ll J$), where thermal monopoles (ice-rule violations) are strongly suppressed, this same transition into the $\bm{q}=0$ ice-ferro state can be viewed as a topological transition driven by the attraction between extended one-dimensional Dirac strings (system-spanning string excitations), as established for spin ice with further-neighbor exchange~\cite{Chou2010, Jaubert2021}. Because a Dirac string carries an extensive energy cost set by its line tension, a finite temperature is required to nucleate a single string; once nucleated, however, the effective attraction between strings triggers their proliferation through a first-order transition. This string-mediated picture is fully consistent with our finite-temperature results: the line tension of these one-dimensional defects allows a genuine (first-order) transition to survive, whereas the proliferation of point-like thermal monopoles---which act as endpoints that sever the strings~\cite{Watanabe2026}---is precisely what reduces the ice-VII--ice-X pathway to a continuous crossover. Thus, while the thermodynamic boundary of the ice-VIII phase is protected by the spontaneous breaking of macroscopic spatial symmetries, the pathway between ice-VII and ice-X lacks such spatial symmetry breaking. Consequently, it is governed by the topological fragility of the emergent gauge field against thermal point-like defects, making a continuous crossover inevitable.

The implications of our macroscopic symmetry perspective may extend beyond the ultrahigh-pressure phases. For instance, the proton-disordered ice-III and its low-temperature counterpart ice-IX share the identical macroscopic space group ($P4_12_12$)~\cite{RevModPhys.84.885, Salzmann2011}. Indeed, neutron diffraction studies have revealed that ice-III is already partially ordered while ice-IX retains residual disorder, suggesting an evolution of the degree of order rather than a complete symmetry breaking~\cite{Lobban2000}. Analogous to the relationship between ice-VII and ice-X, the absence of macroscopic spatial symmetry breaking raises a fundamental question of whether ice-III and ice-IX are separated by a genuine thermodynamic singularity or connected via a continuous crossover. Exploring whether the topological fragility of the hydrogen-bond network also governs the phase behaviors of these lower-pressure polymorphs remains a fascinating direction for future study.

In conclusion, our study confirms that within the effective spin model, the putative topological phase boundaries between the molecular and non-molecular phases are smeared into continuous crossovers by monopole screening. Our microscopic numerical results, consistent with recent analytical derivations~\cite{Watanabe2026} and macroscopic thermodynamic models~\cite{Ackland2025}, verify that ice-VII and ice-X are different regimes of the same thermodynamically indistinguishable phase at finite temperatures. The distinction between them is marked not by a phase boundary, but by the continuous evolution of proton dynamics, consistent with the supercritical crossover pathways observed in experiments.
We stress that the strength of such a minimal model lies not in reproducing every structural detail of dense ice, but in isolating the mechanism that governs its thermodynamics. The model retains the ingredients essential to ice-VII, VIII, and X---the Bernal-Fowler ice-rule manifold and its monopole excitations, the degeneracy-lifting that selects the proton-ordered ice-VIII state, and the single-ion term that drives bond symmetrization into ice-X---and from these alone it reproduces the topology of the high-pressure phase diagram and resolves the thermodynamic status of the ice-VII--ice-X transformation.

\begin{acknowledgments}
We thank Zhaoyu Han, Kazuki Komatsu, Naoto Nagaosa, Tibor Rakovszky, Inti Sodemann, Masaki Oshikawa, Ryan Thorngren, and Han Yan for useful discussions.
The work of S.W. is supported by JST SPRING, Grant No.~JPMJSP2108.
The work of Y.M. is supported by JSPS KAKENHI Grant No.~JP25H01247.
The work of H.W. is supported by JSPS KAKENHI Grant No.~JP24K00541.
The work of H.W. was conducted in part during the program ``Generalised symmetries and anomalies in quantum phases of matter 2026'' (code: ICTS/GSYQM2026/01).
\end{acknowledgments}

\bibliography{ref_ice}

\end{document}